\documentclass[prl,twocolumn,superscriptaddress,showpacs,epsf]{revtex4}

\usepackage{graphicx}
\usepackage{latexsym}
\usepackage{amssymb}
\usepackage{amsfonts}
\usepackage{soul}
\usepackage{color}
\begin{document}
\bibliographystyle{apsrev}
\title{Tunable Field Induced Superconductivity}

\author{W. Gillijns, A.V. Silhanek, and V. V. Moshchalkov}

\affiliation{INPAC - Institute for Nanoscale Physics and
Chemistry, K.U.Leuven, Celestijnenlaan 200D, B--3001 Leuven,
Belgium}


\date{\today}
\begin{abstract}
We investigate the transport properties of a thin superconducting
Al layer covering a square array of magnetic dots with
out-of-plane magnetization. A thorough characterization of the
magnetic properties of the dots allowed us to fine-tune their
magnetic state at will, hereby changing the influence of the dots
on the superconductor in a continuous way. We show that even
though the number of vortex-antivortex pairs discretely increases
with increasing the magnetization of the dots, no corresponding
discontinuity is observed in the resistance of the sample. The
evolution of the superconducting phase boundary as the magnetic
state of the dots is swept permits one to devise a fully
controllable and erasable field induced superconductor.
\end{abstract}

\pacs{74.78.-w 74.78.Fk 74.25.Dw}

\maketitle

The hallmark of superconductivity and the technological
applications based on it is the possibility to carry electrical
currents without resistance below the critical temperature
$T_{c}$. This non-dissipative quantum state is however suppressed
either by applying a magnetic field or by submitting the system to
a high enough electrical current as to break the Cooper-pairs
responsible for the superconducting condensate. In Type-II
superconductors which are most interesting for practical
applications, an applied magnetic field $H$ ($H_{c1} < H <
H_{c2}$, with $H_{c1}$ and $H_{c2}$ the lower and upper critical
fields) penetrates the superconductor in the form of flux quanta.
Under the influence of an applied current these vortices start
moving, hereby destroying the non-dissipative state. During the
last decade enormous efforts have been devoted to prevent this
drawback by anchoring vortices with different types of pinning
potentials. Particular attention has been focused on magnetic
pinning centra originally due to their promising enhancement of
the critical current \cite{Bulaevskii-2000}. Interestingly, it was
also found that in such superconductor/ferromagnet (S/F) hybrid
systems field compensation effects between an applied magnetic
field and the stray fields of the ferromagnets can drastically
change the superconducting properties \cite{Buzdin-2005,
Lyuksyutov-2005}.

It has recently been shown that if an array of out-of-plane
magnetized dots is deposited on top of a superconducting film, the
$T_{c}(H)$ phase boundary can be \emph{shifted by exactly an
integer number of flux quanta per unit cell} \cite{Lange-2003a,
Milosevic-2004}. The occurrence of a maximum in $T_{c}$ at a
non-zero magnetic field results from the compensation of the dots'
field by the applied field. The reason for the quantized character
of field-induced superconductivity (FIS) lies in the ability of
the superconductor to quantize the flux generated by a magnetic
dot, whatever its value, by either compensating the field excess
or generating the field deficiency with supercurrents. If the dots
generate a small flux, supercurrents will counterbalance this flux
and no compensation effects are present. However, if the flux
crosses a certain critical value the field lines generated by the
dot penetrate the superconductor in the form of vortex-antivortex
(V-AV) pair(s) and hence and the maximum critical temperature will
shift by exactly an integer number of flux quanta per unit cell.
This picture suggests that for particular magnetization values the
field locus of the maximum $T_{c}$ should undergo abrupt
displacements from $n\phi_{0}$ to $(n+1)\phi_{0}$ with $n$
integer. So far, the detailed evolution of this shift by an
integer number of flux quanta has remained unveiled, mostly due to
the difficulties to control the magnetization of the dots in a
continuous fashion. Here we explore the evolution of the
superconducting properties of a thin film superconductor, now
deposited on top of an array of \emph{tunable} magnetic dots.

The sample under investigation is a superconducting Al film of
thickness $d = 50$ nm evaporated on top of a square array of
magnetic dots with 2 $\mu$m lattice constant. A 5 nm thick Si
buffer layer was first evaporated on top of the dots to avoid
proximity effects. In this way the interaction between the
ferromagnet and the superconductor is purely electromagnetic in
origin. An Atomic Force Microscopy picture of the dots' array is
shown in Fig. \ref{Fig1}. The ferromagnetic dots have a diameter
of 1.36 $\mu$m and consist of a 2.5 nm Pt buffer layer covered
with a [0.4 nm Co/1.0 nm Pt]$_{10}$ multilayer with magnetization
perpendicular to the sample surface \cite{Zeper-89}.

The magnetic properties of the dots were investigated using a
commercial Quantum Design SQUID magnetometer. The main panel of
Fig. \ref{Fig1} presents the field dependence of the magnetization
$M(H)$ for the Co/Pt dots at $T$=5 K, where the magnetization was
determined by using the total volume of the dot. The broader magnetization loop (filled circles)
was recorded after saturation in a field of $\mu_{0}H_{s}$=1 T.
After following a careful demagnetization procedure it is possible
to reduce the remanent magnetization down to 0.5\% of the
saturation value. As a consequence of the large diameter of the
dots, the demagnetized state microscopically corresponds to a
magnetic multidomain state with very little stray field
\cite{Lange-2003a}. Starting from the demagnetized state minor
loops can be built up by making field excursions $H_{m}$ with
$H_{m} < H_{s}$ (open symbols). Every minor loop has a unique
remanent magnetization value $M_{0}$ associated with it. Repeating
this procedure for several $H_{m}$ we determined $M_{0}$ as a
function of $H_{m}$ as shown in Fig. \ref{Fig1}(b). It is
important to note that the full control of the out-of-plane
magnetization of the dots relies on the one to one correlation
between $M_{0}$ and $H_{m}$ together with the reproducibility of
the minor loops.\cite{comment}

From the normal / superconductor (N/SC) phase boundary for the
demagnetized dots we determine a critical temperature $T_{c}$ of
\hbox{1.343 K} and a superconducting coherence length
\hbox{$\xi(0)$ = 117 nm}. Using the electronic mean free path $l
\approx 15 \pm 1$ nm $<< \xi(0)$ as estimated from the normal
state resistance above $T_{c}$ we determined a penetration depth
$\lambda(0) \sim$ 100 nm in the dirty limit\cite{Thinkham}. The effective penetration depth due to the
thin film geometry is $\Lambda = \lambda^2/d$, which gives 200 nm
for our sample. From this values we obtained a Ginzburg-Landau parameter
$\kappa=\Lambda/\xi$ $\approx$ 1.7 showing that our sample is a Type-II
superconductor.

Let us first analyze the evolution of the N/SC phase boundary
$T_{c}(H)$ as the magnetization of the dots is gradually
increased. Figure \ref{Fig2}(a) shows a selected set of these data
determined by a 10\% normal state resistance criterion. As
expected, when the dots are in the demagnetized state ($M_{0}$=0)
a symmetric $T_{c}(H)$ boundary is observed. It is worth noticing
that this boundary nearly reproduces that corresponding to the
virgin state (dotted line) as a consequence of having a similar
multidomain state of the dots. As the out-of-plane magnetization
of the dots crosses some critical value, a vortex is created on
top of the dot while the corresponding antivortex is located in
between the dots. The presence of this vortex antivortex pair (V-AV) shifts $T_{c}^{max}$ to
$\phi$ = $\phi_{0}$. The reason for this shift is that an external
applied field $\phi$ = $\phi_{0}$ will introduce an extra vortex
per unit cell which in turn annihilates the interstitial
antivortex. Since the effective field between the dots is minimal
for this particular field a maximum $T_{c}$ is obtained in this
case. Further increase of the magnetization of the dots results in
a shift of the phase boundary by an integer number $n$ of quantum
flux units $\phi_{0}$ and in a decrease of the maximum
$T_{c}^{max}$. This decrease of $T_{c}^{max}$ is a result of the
increasing average field felt by the superconductor. It is clear
that by controlling the magnetic states of the dots it is possible
to place the position of $T_{c}^{max}$ at any desired $n<$ 7. The
maximum shift is ultimately determined by the maximal flux
generated by the dots. This value can be increased by either
increasing the dot size or by increasing the saturation
magnetization.

For all $M_{0}$ values and sufficiently high temperatures it can
be seen that $T_{c}(H)$ exhibits a parabolic background. A similar
behavior has been reported for F/S bilayers \cite{Lange-2003b} and
samples with square arrays of antidots \cite{Rosseel-1997} and can be attributed to the change of dimensionally when $\xi(T)$
exceeds the width $w$ of the areas where superconductivity first
nucleates. Within this regime the phase boundary can be
approximated by\cite{parabolic} $T_{c}(H)/T_{c}(0)=1-(\alpha H)^2$, with $\alpha =
\xi(0)\pi w/2 \sqrt 3\phi_{0}$. By using this expression to fit
the data in Fig. \ref{Fig2}(a) we estimate $w$ as a function of
$M_{0}$ (as shown in Fig. \ref{Fig2}(b)). Here a continuous
decrease of $w$ with $M_{0}$ is observed which is consistent with
the reduction of the available nucleation area due to the
increasing strength of the magnetic template created by the dots.

The previous description of the evolution of $T_{c}(H)$ with
increasing $M_{0}$ would remain incomplete without determining
whether the transition from $n$ to $n+1$ V-AV pairs actually
occurs as a sudden horizontal displacement. In order to address
this issue we monitored the evolution of the resistance $R(H)$ at
a nearly constant reduced temperature $t$ = 0.991 as the
magnetization of the dots $M_{0}$ is increased. This procedure
turns out to be far more sensitive than just following the
evolution of the $T_{c}(H)$ phase boundary itself. In Fig.
\ref{Fig3}(a) we present these measurements for the particular
case of $n$=1 although the complete spectrum of accessible
magnetization values was experimentally spanned \cite{Movie}. For
the sake of clarity the curves in Fig. \ref{Fig3}(a) have been
displaced horizontally along the field axis.

The leftmost curve exhibits a very symmetric shape with a minimum
resistance centered at $\phi=\phi_{0}$ and two local dips at
$\phi=0$ and $\phi=2\phi_{0}$. This particular magnetization value
corresponds to the generation of exactly one V-AV pair. As $M_{0}$
is increased the central dip slowly moves upward whereas the
satellite dip at $\phi =2\phi_{0}$ becomes deeper. Strikingly, at
a certain critical magnetization $M_{c2}$ both local minima reach
the same level. From that point on the absolute minimum resistance
is located at $\phi=2\phi_{0}$ and eventually a new symmetric
configuration centered at  $\phi=2\phi_{0}$ is obtained. The above
described evolution shows that even though the absolute minimum of
the $R(H)$ jumps from $\phi=\phi_{0}$ to $\phi=2\phi_{0}$ at
$M=M_{c2}$, at any field the resistance remains a continuous
function of $M_{0}$. A similar evolution is observed for switching
from $n$ to $n+1$ with $n<$7 \cite{Movie}.

In order to explain the above described behavior it is crucial to
take the screening supercurrents $J$ into consideration, very much
like in the Little-Parks effect \cite{Little-Parks} (see Fig.
\ref{Fig3}(b)). For $M_{0}$ = 1.67 kA/cm, at $\phi=\phi_{0}$ one
flux line is located at the magnetic dot and no total currents are
present since the flux of the vortex is provided by the field of
the dot (black circle in Fig. \ref{Fig3}(b)). Under these
circumstances both vortices and anti-vortices feel the same
interaction with the dot and thus the satellite dips at $\phi=0$
and $\phi=2\phi_{0}$ are equally deep. Upon increasing the
magnetization, the excess of field generated by the dot is
counteracted by the supercurrents (blue and red circles in Fig.
\ref{Fig3}(b)). These currents break the previous symmetry
favoring the presence of vortices over anti-vortices thus
accounting for the deepening of the minimum at $\phi=2\phi_{0}$
and the rising of the minimum at $\phi=0$. This unbalanced
situation persists up to a certain magnetization $M_{c2}$.
Crossing this value (green circle in Fig. \ref{Fig3}(b)) results
in the creation of an extra V-AV pair, thus shifting the phase
boundary to $\phi=2\phi_{0}$. At the same time the currents
circling the dots have reversed polarity, now favoring
antivortices rather than vortices. This interaction is reflected
in the reversal of the asymmetry with respect to the new minimum.
Further increasing $M_{0}$ reduces the asymmetry as the current
progressively approaches zero (cyan and magenta circles in Fig.
\ref{Fig3}(b)). Eventually a fully symmetric curve centered at
$\phi=2\phi_{0}$ is recovered (gold circle in Fig. \ref{Fig3}(b)).

It is worth emphasizing that although the local minima in the
resistance move up and down in a continuous manner as $M_{0}$ is
swept, the field position of the absolute minimum of the $R(H)$
curves always undergoes a discrete jump every time a critical
value $M_{c}$ is crossed. This situation becomes more evident in
Fig. \ref{Fig4} where a contour plot of $R(H,M_{0})$ at $t$ =
0.991 is shown. The red islands in this graph indicate the
location of the lower resistance. A clear stepwise structure
associated with the discrete increase of V-AV pairs as $M_{0}$
rises, can be discerned. From this plot one can accurately
determine the magnetization needed to generate the first, second,
third, and fourth V-AV pairs.

It is interesting to note that the generation of the first
vortex-antivortex pair appears delayed with respect to the
subsequent steps. This finding is consistent with previous
theoretical predictions by Milosevic et al. \cite{Milosevic-2004}
based on the solution of the Ginzburg-Landau equations. In that
work it is shown that for an analogous hybrid system with similar
superconducting and magnetic properties to ours the induction of
the first V-AV pair needs a larger flux than for the following
transitions. The ultimate reason for this effect lies in the
higher degree of symmetry of the $n$=0 V-AV state with respect to
the $n\neq$ 0 states. Our results represent the first experimental
confirmation of this prediction. In a later report the same
authors \cite{Milosevic-2005} have predicted that the
$T_{c}(M_{0})$ phase boundary corresponding to an array of tunable
magnetic dots should exhibit cusp-like features associated with to
the generation of V-AV pairs superimposed on a global decrease of
$T_{c}$ with increasing $M_{0}$. This global reduction of $T_{c}$
is clearly visible in Fig. \ref{Fig2}(a). And since the critical
temperature is determined by the lowest resistance, the crossing
of the two minimal dips in the $R(H)$ curves at $M_{c}$ will
result in a cusp-like feature in the $T_{c}(M_{0})$ phase boundary
as well.

As a last remark, we would like to point out that the observed
similarities with the Little-Parks \cite{Little-Parks} effect
should not be pushed too far. Although in both cases the currents
adjust themselves to quantize the magnetic field, in our system
the field is provided by the dots rather than by a homogeneous
source giving rise to V-AV pairs.  Also, the spatial coexistence
of screening and vortex currents is not present in our system
since vortices can move around in the two-dimensional lattice.

In conclusion, we have shown that the remanent magnetization
$M_{0}$ of out-of-plane magnetized dots can be tuned from nearly
zero (demagnetized) to a material dependent maximum value. This
continuous degree of freedom allowed us to observe and investigate
\emph{tunable field induced superconductivity} and the progressive
evolution of the superconducting phase boundary as a function of
$M_{0}$ and thus reveal the microscopic mechanisms that lead to
discrete jumps in $T_{c}^{max}$ without having discontinuities in
the resistance at any point. The remarkable flexibility attainable
with these magnetic field resistant superconductors makes them
very attractive for practical applications.

This work was supported by the Fund for Scientific
Research-Flanders FWO-Vlaanderen, the Belgian Inter-University
Attraction Poles IUAP, the Research Fund K.U. Leuven GOA/2004/02
and by the Japan-EU-USA Nanoscience and Engineering in
Superconductivity (NES) programs. A.V.S. is grateful for the
support from the FWO-Vlaanderen. We would also like to thank IMEC
for making the resist patterns.

\begin{figure}[b!]
\caption{(color online) Magnetization loops of the Co/Pt dots for
three different excursion fields measured at 5 K. The insets show
(a) an Atomic Force Microscopy-picture of the magnetic dots, and
(b) the remanent magnetization $M_{0}$ as a function of the
maximum applied field $H_{m}$ starting from the demagnetized state
$\mu_{0}H_{m}$ = \hbox{0 mT}.} \label{Fig1}
\end{figure}

\begin{figure}[b]
\caption{(color online) (a) Superconducting transition $T_{c}(H)$
of the Al film for different magnetic states of the dots. By
increasing the magnetization a clear shift of $T_{c}(H)$ and a
decrease of $T_{c}^{max}$ is observed. (b) Lateral dimension $w$
of the nucleation of superconductivity as a function of the
magnetization of the dots.} \label{Fig2}
\end{figure}

\begin{figure}[t]
\centering
\caption{(color online) (a) Magnetoresistance measurements for
different values of the magnetization $M_{0}$ of the magnetic dots
at a constant reduced temperature $t = T/T_{c}$. The presented
data corresponds to the transition of the location of the maximum
$T_{c}$ from $\phi=\phi_{0}$ to $2\phi_{0}$. For clarity the
curves have been displaced horizontally. (b) Illustration of the
supercurrents encircling the magnetic dots as a function of the
magnetization. Colored symbols correspond to the colored curves in
panel (a). Starting from the black curve, the increase of the
magnetization will result in increasing currents, counteracting
the field of the dots and keeping the flux exactly equal to
$\phi_{0}$. Crossing a critical value $M_{c2}$ the currents
reverse polarity hereby generating an extra vortex-antivortex
pair. Further increasing the magnetization will result in a
decrease of the currents since the flux is increasingly carried by
the dot itself.} \label{Fig3}
\end{figure}

\begin{figure}[h!]
\caption{Contour plot of the resistance as a
function of field and magnetization for $t$ = 0.991. Red islands
correspond to lowest resistance and hence highest critical
temperature. The step-like profile is a consequence of the
quantization of the magnetic field in an integer number of flux
quanta by the superconductor.} \label{Fig4}
\end{figure}


\begin{references}
\bibitem{comment}
It is worth noticing that the magnetization of the dots exhibit almost no temperature dependence for 5 K $<$ T $<$ 300 K. Due to the large separation between dots there is little interaction between them and the properties of the array mimic the properties of an individual dot. Although the present report deals with circular shaped dots, both the tuneability and demagnetization procedure are independent of the actual geometry of the dots.
\end{references}
\end{document}